# Indirect observation of molecular disassociation in solid benzene at low temperatures

F. Yen[1,2*], S. Z. Huang[1], S. X. Hu[1], L. Y. Zhang[1], L. Chen[1†]

[1]Department of Physics, Southern University of Science and Technology of China, 1088 Xueyuan Blvd. Shenzhen, Guangzhou 518055 P. R. China
[2]Key Laboratory of Materials Physics, Institute of Solid State Physics, Hefei Institutes of Physical Science, Chinese Academy of Sciences, Hefei 230031, P. R. China

*Fei Yen, fyen@sustc.edu.cn; †Lang Chen, chenlang@sustc.edu.cn

**Abstract:** The molecular dynamics of solid benzene are extremely complex; especially below 77 K, its inner mechanics remain mostly unexplored. Benzene is also a prototypical molecular crystal that becomes energetically frustrated at low temperatures and usually unusual phenomena accompanies such scenarios. We performed dielectric constant measurements on solid benzene down to 5 K and observed a previously unidentified minimum in the imaginary part of the dielectric constant at $T_m$=17.9 K. Results obtained on deuterated solid benzene ($C_6D_6$, where D is deuterium) show an isotope effect in the form of a shift of the critical temperature to $T_{m'}$=18.9 K. Our findings indicate that at $T_m$, only the protons without the carbon atoms continue again to undergo rotational tunneling about the hexad axes. The deuterons appear to do the same accounting for an indirect observation of a continued and sustained 12-body tunneling event. We discuss how similar experiments performed on hydrogen-based molecular crystals can be exploited to help us obtain more insight on the quantum mechanics of many-body tunneling.

Benzene is one of the most prototypical molecules in organic chemistry: it is the elemental unit of all aromatic compounds and the π-bonding of its carbon atoms serves as the quintessential example of electron delocalization. The molecular dynamics of the $C_6H_6$ molecule are extremely complex and not yet fully understood; only until recently have the inner dynamics of the benzene dimer been unraveled [1]. In the solid phase, the $C_6H_6$ molecules, or at least a subset of them, undergo rotational motion about their $C_6$ hexad axes starting from just below the melting point [2,3]. Periodic oscillations about its $C_2$ diagonal axes also occur, albeit at lower frequencies [4,5]. As the system is further cooled, rotational motion slows down, particularly upon approaching near 230 K, the angular speeds experience an abrupt decrease due to an increase of the periodic potential barrier heights [6,7]. Near 120 K, nuclear magnetic resonance measurements indicate that reorientations of the $C_6H_6$ molecules take place [8-10]. It is expected that at some point in the low temperature regime, molecular motion must seize, at least in the classical sense, since the heights of the periodic potential barriers continue to elevate with decreasing temperature. However, not much is known pertaining the molecular dynamics of solid benzene starting from below 77 K. Usually, many interesting fundamental processes can be observed at low temperatures due to the absence of thermal noise.

Benzene crystallizes into the orthorhombic structure space group *Pbca* just below 279 K (Fig. 1a) [11]. Solid benzene can be regarded as yet another type of prototype; that of a molecular crystal as there is a clear distinction between its inter- and intra-molecular bonds [12]. The unit cell possesses four $C_6H_6$ molecules with the centers of different molecules situated at each of its eight corners and at the center of each of its six faces [13]. According to calculations of the Gibbs free energy, a phase transition is expected to take place upon cooling below around 100 K from the *Pbca* to the *Cmca* representation [14] (Fig. 1b), also orthorhombic in structure. One key difference is that *Cmca* only possesses two types of molecules instead of four in *Pbca*. Hence, during the phase transition two out of the four sets of molecules must undergo a near 90° reorientation. The glass transition temperature $T_g$ of benzene is believed to reside above 140 K according to studies of its confinement in small cavities [15] and mixtures with other highly vitreous liquids [16]. Usually in such cases where a structural phase transition rests below the $T_g$ of the system, a complete transition never takes place because ionic motion is prohibited; even proton hopping is deemed to be forbidden. For instance, hexagonal ice is unable to phase transition into its ground state, which occurs at 72 K [17] because its $T_g$ rests at 136 K [18]. This begs the question as to whether the *Cmca* phase in solid benzene is accessible or not. If not, what interesting phenomena accompany such an energetically frustrated system? Continuous measurement of the complex dielectric constant yields information on any changes in the charge distribution as well as its *rate* of change for which in this case, is pertinent for studying the molecular dynamics of benzene. Moreover, there is no existing literature on the experimental values of the dielectric constant of solid benzene toward low temperatures.

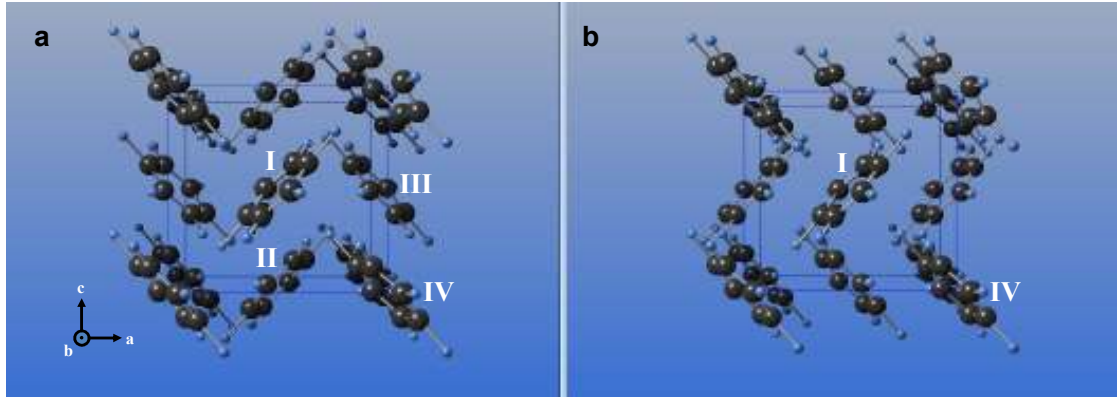

**FIG. 1** (color online). Crystalline benzene in the Space groups (a) *Pbca* and (b) *Cmca*. In the *Pbca* structure, four types of molecules exist: three situated at the centers of the faces of the unit cell labeled as I, II, & III; and one type at the corners as IV. The *Cmca* representation has only two types of molecules, I and III.

Liquid benzene $C_6H_6$ acquired from Aladdin Industrial Corp., Ltd. Shanghai (99.95% purity) was injected into a cylindrical Teflon (PTFE) container 3 mm in radius and 12 mm in height. An oxygen-free copper cap equipped with two platinum electrodes in the form of a pair of parallel plates was used to seal the sample container in an argon atmosphere. More details can be found elsewhere [19]. Polycrystalline benzene was formed by cooling the sample at 2-3 K/min past its melting temperature where it remained in supercooled form down to 249 K where crystallization occurred marked by the decrease in the real part of the dielectric constant $\varepsilon'(T)$ by nearly 60% (Fig. S1). Upon warming, a sharp transition was observed in $\varepsilon'(T)$ at $T_M=278.66$ K, only 0.02 K away from the melting point of benzene. The electrodes were connected to an Andeen Hagerling capacitance bridge (model AH2700A). The values of $\varepsilon'(T)$ and the imaginary part of the dielectric constant $\varepsilon''(T)$ were derived from the capacitance and loss tangent of the bridge, respectively. The temperature environment was controlled by a PPMS cryostat made by Quantum Design, Inc. For the measurements on deuterated solid benzene $C_6D_6$ (where D is deuterium), the same preparation and measurement methods were employed, but instead, liquid $C_6D_6$ from Aladdin Industrial Corp. Ltd., Shanghai (99.6% purity) was used.

Figure 2 shows the real $\varepsilon'(T)$ and imaginary $\varepsilon''(T)$ parts of the dielectric constant with respect to temperature for $C_6H_6$ and $C_6D_6$ at 1 kHz. The most pronounced feature are the minima at $T_m=17.9$ K and $T_{m'}=18.9$ K for $C_6H_6$ and $C_6D_6$, respectively. In addition, a small shoulder is identified at $T_s=32$ K for $C_6H_6$ and $T_{s'}=33$ K for $C_6D_6$. Dielectric dispersion $\varepsilon'(\omega)$ and absorption $\varepsilon''(\omega)$ measurements were obtained at different temperatures and are shown in the supplementary information section as Figs. S2 and S3. Figure 3 plots $\varepsilon'(\omega)$ and $\varepsilon''(\omega)$ at their matched frequencies from which $T_m$ can be observed to remain nearly independent of frequency up to the limit of our high-precision capacitance bridge of 20 kHz.

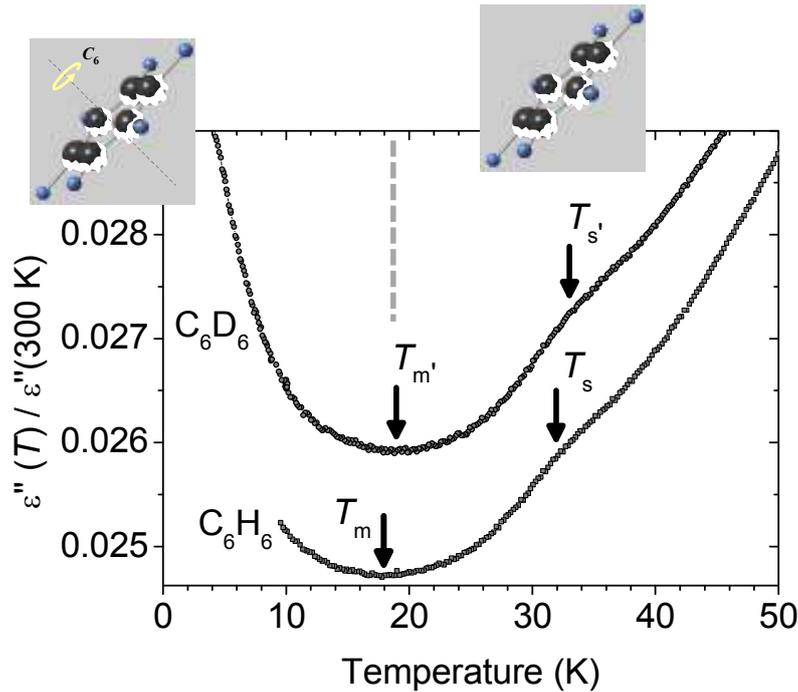

**FIG. 2** (color online). Temperature dependence of the dielectric constant $\varepsilon''(T)$ at 1 kHz. The minima at $T_m$=17.9 K and $T_{m'}$=18.9 K indicate that the protons and deuterons, respectively, undergo rotational tunneling in $C_6H_6$ and $C_6D_6$ in the *Pcba* phase. $T_s$=32 K and $T_{s'}$=33 K.

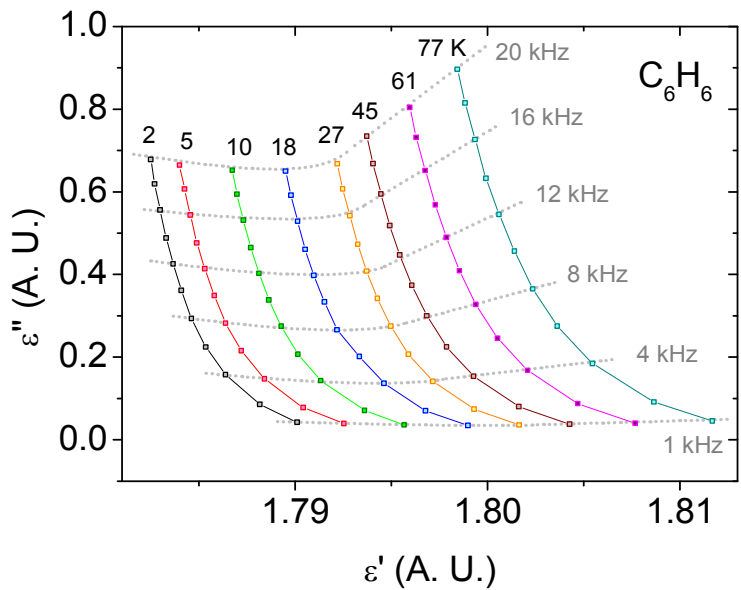

**FIG. 3** (color online). Dielectric dispersion and absorption plotted at their matched frequencies linked by hand-drawn dotted lines. $T_m$ is present in $\varepsilon''(T)$ up to 20 kHz, the experimental limitation of our equipment.

At higher temperature, a step discontinuity in $\varepsilon''(T)$ was observed near $T_c$=227 K (Fig. 4) which is in excellent agreement with interpretations on nuclear magnetic resonance data by Das [6] concluding that a reduction of the periodic potential barrier occurs at temperatures above 230 K. Craven *et al.*, also observed a marked increase of the $C_1$-$D_1$ bond angle and the *b* lattice parameter above 230 K [13,20]. The inset of Fig. 4 shows the first derivative of $\varepsilon''(T)$ with respect to temperature where a maximum and minimum are observed at 208 and 238 K, respectively, which require 30 K to complete. The sharp discontinuities suggest that perhaps two successive first order phase transitions occur rather than one that is second order in nature. Also in $d\varepsilon''(T)/dT$, a change in slope discontinuity is observed at $T_R$=126 K which coincides well with the critical temperature reported previously regarding reorientations of the benzene molecules starting from above 120 K [8-10].

The melting point of benzene is 1.3 K lower than that of its deuterated analogue. The observed shifts in $T_m$ and $T_s$ by nearly the same values suggest that they are intrinsic of benzene and are not likely due to vacancy diffusion or motion of its π-electrons. In contrast to $\varepsilon''(T)$, $\varepsilon'(T)$ did not exhibit any type of discontinuity in the same temperature range. Usually, $\varepsilon''(T)$ is proportional to the first derivative of $\varepsilon'(T)$. Moreover, as temperature is decreased toward absolute zero, $\varepsilon''(T)$ is supposed to decrease to a finite value due to a zero point energy associated to the system. The unusual increase in $\varepsilon''(T)$ while $\varepsilon'(T)$ remains monotonic means that the local quadrapole moment distribution remained unchanged; however, components of the distribution experienced motion, viz. an increase of the ground state energy. From a different perspective, $\varepsilon''(T)$ is proportional to the real part of the conductivity. One explanation for the increase in charge mobility below $T_m$, which is highly unusual in this case, is that two or more ions switched positions in a concerted manner since all of the electrons are in the insulating band. This fits the idea that the molecular motions occur at 60° increments as suggested in several related literature [13] since after each event the local quadrapole moment distribution is unaffected even when the rate of rotation changes. If indeed molecular motion takes place, then which ions are involved below $T_m$?

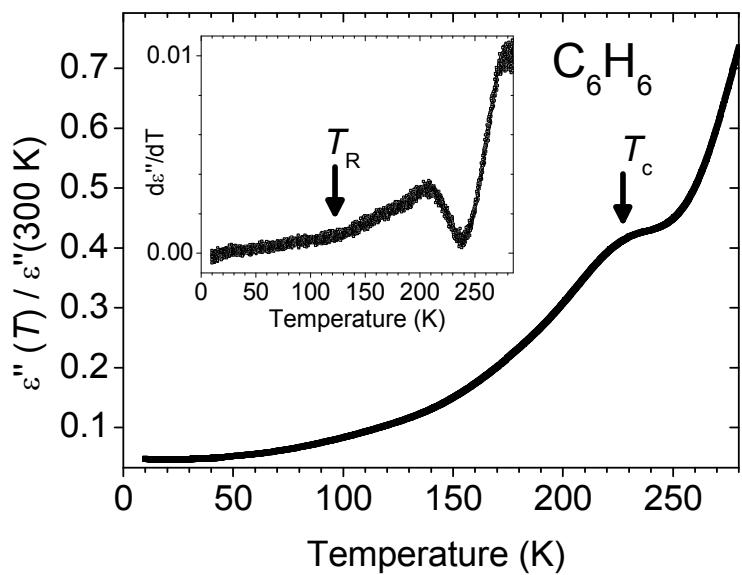

**FIG. 4**. Imaginary part of the dielectric constant at high temperatures at 1 kHz. $T_c \approx 235$ K is in agreement with existing literature where the rates of the periodic potential barrier [6] and b-axis experience a change [13,20]. Inset: $T_R \approx 126$ K coincides with the temperature reported by Andrew *et al.* attributed to reorientations of the benzene molecules [8-10].

Below $T_g$, ionic motion can only occur via quantum tunneling. It is therefore expected that the rotational motion of the benzene molecules must come to an arrest just below $T_g$. This explains why the *Cmca* ground state is inaccessible; our $\varepsilon'(T)$ indicate no structural phase transitions and no experimental studies have reported the observation of the *Cmca* phase. For instance, Craven *et al.* emphasize that no phase transitions occur down to 4 K from their neutron powder diffraction experiments [13]. Recently, quasi-static rotational motion has been found to be the means for a molecule to steer itself toward the energy landscape with the lowest potential barrier [21,22]. As such, the system becomes energetically frustrated at low temperatures where the molecules have the tendency to continue to rotate but are constrained to do so. With decreasing temperature, the height of the potential barriers increase, but they also become narrower because of thermal expansion. The C-H and H-H distances are affected the most since C-C covalent bonds are usually independent of temperature in organic compounds. Moreover, the unusual narrowing of the benzene C-H overtone at 1.8 K to values equivalent to its gas phase suggests that displacements of the C-H bonds take place at such temperatures [23]. Hence, we speculate that near $T_m$, the potential barriers become narrow enough that the protons are capable to tunnel so the $C_6$ rotations continue, albeit sans the carbon atoms as the latter are too massive to tunnel. Similar phenomenon has been directly visualized in ice where correlated tunneling of protons occurs starting from just below 20 K [24]. Also, the same dielectric constant signature was observed in ice [25,26] where there is a general consensus that protons undergo tunneling in a six-body coordinated manner to neighboring empty sites within $H_2O$ hexamers [27-29]. However, after the tunneling process the system remains equally disordered and equally frustrated due to the six-fold symmetry of the lattice. As such, the protons proceed to tunnel back to their original sites and the process is repeated over and over. As the temperature decreases, the population of protons undergoing correlated tunneling increases which explains the increase of $\varepsilon''(T)$ starting from $T_m$. Hence, we conclude that $T_m$ marks the commencement of correlated tunneling of protons (deuterons) in groups of six about the hexad axes of the $C_6H_6$ ($C_6D_6$) molecules.

Whether the concerted decoupling of the six protons from the six carbon atoms occur instantaneous is a subject beyond our scope. However, our findings indicate that this process may be a possible elementary step of the reaction mechanism in non-polar molecules. Equally important is that this is the first time where deuterons are indirectly observed to tunnel in what can be deemed as a continued and sustained 12-body (six protons plus six neutrons) tunneling event. The tunneling process of one single electron is readily understood. For the case of two-body tunneling our knowledge becomes more limited such as Cooper pairs traversing through Josephson junctions [30] and tautomerization of protons in DNA strands [31]. As for multi-body tunneling, we are merely at the tip of the iceberg. The presence of hydrogen bonds translates to low potential energy barriers (or shallow valleys) and easily excitable states which both increase the tunneling transmission coefficient. When a structural phase transition lies below the glass transition temperature the protons are induced to play musical chairs with its next nearest neighbors or tunnel to unoccupied adjacent sites but in either case are constrained to do so in groups in order to avoid energy penalties. This renders hydrogen-based molecular crystals a nonconventional platform for studying the thermodynamics of molecular bosons and probing the quantum mechanics of many-body tunneling at the macroscopic level.

**Acknowledgements:**

This work was made possible by: the National Natural Science Foundation of China, grant numbers 11374307, 11250110050, 11474146，11504161, 11650110430 and U1532142; the Hong Kong, Macao and Taiwan Science & Technology Cooperation Program of China grant No. 2015DFH10200; the Science and Technology Research Items of Shenzhen grant No. JCYJ20140417105742706; and a Nanshan Key Laboratory support program.